\documentclass[apj]{emulateapj}
\usepackage{apjfonts} 

\usepackage{graphicx}

\makeindex
\citeindextrue

\received{2007, July~2}
\revised{2007, August~16}
\accepted{2007, August~20}
\journalinfo{Astrophysical~journal Letters}
\submitted{}

\shorttitle{The Inner Jet of M87}
\shortauthors{Kovalev et al.}

\begin{document}
\title{The Inner Jet of the Radio Galaxy \objectname{M87}}
\author{Y. Y. Kovalev\altaffilmark{1,2},
        M. L. Lister\altaffilmark{3},
        D. C. Homan\altaffilmark{4},
        K. I. Kellermann\altaffilmark{5}
}

\altaffiltext{1}{Max-Planck-Institut f\"ur Radioastronomie,
                 Auf dem H\"ugel 69, 53121 Bonn, Germany; 
                 \mbox{ykovalev@mpifr-bonn.mpg.de}
}

\altaffiltext{2}{Astro Space Center of Lebedev Physical Institute,
                 Profsoyuznaya 84/32, 117997 Moscow, Russia
}

\altaffiltext{3}{Department of Physics, Purdue University,
                 525 Northwestern Avenue, West Lafayette,
                 IN 47907, U.S.A.;
                 \mbox{mlister@physics.purdue.edu}
}

\altaffiltext{4}{Department of Physics and Astronomy, Denison University,
                 Granville, OH 43023, U.S.A.;
                 \mbox{homand@denison.edu}
}

\altaffiltext{5}{National Radio Astronomy Observatory,
                 520 Edgemont Road, Charlottesville,
                 VA~22903--2475, U.S.A.;
                 \mbox{kkellerm@nrao.edu}
}

\begin{abstract}
We report new 2~cm VLBA images of the inner radio jet of
\objectname{M87} showing a limb brightened structure and unambiguous
evidence for a faint 3~mas long counter-feature which also appears limb
brightened. 
Multi-epoch observations of seven separate jet features show typical
speeds of less than a few percent of the speed of light, despite the
highly asymmetric jet structure and the implications of the canonical
relativistic beaming scenario.
The observed morphology is consistent with
a two stream spine-sheath
velocity gradient across the jet, as might be expected from the recently
discovered strong and variable TeV emission as well as from numerical
modeling of relativistic jets.
Considering the large jet to counter-jet flux density ratio and lack of
observed fast motion in the jet, we conclude that either the inner part
of the M87 jet is intrinsically asymmetric or that the bulk plasma flow
speed is much greater than any propagation of shocks or other pattern
motions.
\end{abstract}

\section{INTRODUCTION} 
\label{s:intro}

The peculiar galaxy \objectname{M87} in the \objectname{Virgo} cluster
was among the first to be recognized as a powerful source of radio
emission.  More than 40 years ago, \cite{S64} argued that since radio
galaxies such as \objectname{Cygnus~A} typically have symmetric lobes,
the jets which feed the radio lobes are likely to be intrinsically
two-sided, but that they would appear one-sided due to differential
relativistic Doppler beaming. This simple picture is now widely
accepted as the basis of unified models, which try to explain many of
the different properties of AGN as due to relativistic beaming and
projection effects, rather than to intrinsic differences
\citep[e.g.,][]{UP95}.

M87 remains of great interest today, since there is strong observational
evidence for a $3\times 10^9$ $M_{\sun}$ black hole located at the
galactic nucleus thought to power the relativistic jet \citep{H94,
MMAC97}. Moreover, at a distance of only 16~Mpc, (1~mas = 0.08~pc; 1~mas
per year $=0.25c$) M87 is one of the  closest radio galaxies, and as such
it has one of the few jets which can be well-resolved on sub-parsec
scales in a direction transverse to the flow.

The M87 jet has been studied over a wide range of the electromagnetic
spectrum, including imaging with HST \citep{BSM99} and the VLA
\citep{OHC89} as well as with Chandra \citep{WY02,HCB06}. The observed
morphology at radio, optical, and even X-ray wavelengths appears very
similar, suggesting a common synchrotron radiation mechanism at all
wavelengths and a common spectral index of $-0.67$ throughout the jet
\citep{PBS01}. At least at optical and X-ray wavelengths, the electron
lifetime down the jet is much shorter than the travel time from the
nucleus, suggesting continual in situ acceleration of relativistic
particles within the jet \citep{PBS01}.

In this paper we report on observations made with the NRAO Very Long
Baseline Array (VLBA) at 2~cm wavelength. High dynamic range
images constructed from observations made in the year 2000 describe the
two-dimensional structure of the jet out to nearly 0.2~arcsec (16~pc)
and show the presence of a faint counter-feature. These observations
were complemented by regular observations made with lower sensitivity
between 1995 and 2007 to study the outward flow within the inner part of
the radio jet.

\section{THE OBSERVATIONS AND IMAGING}
\label{s:obs}

\begin{figure*}[t!]
\begin{center}
\resizebox{\hsize}{!}{
   \includegraphics[trim = 0cm 0cm 0cm 0cm]{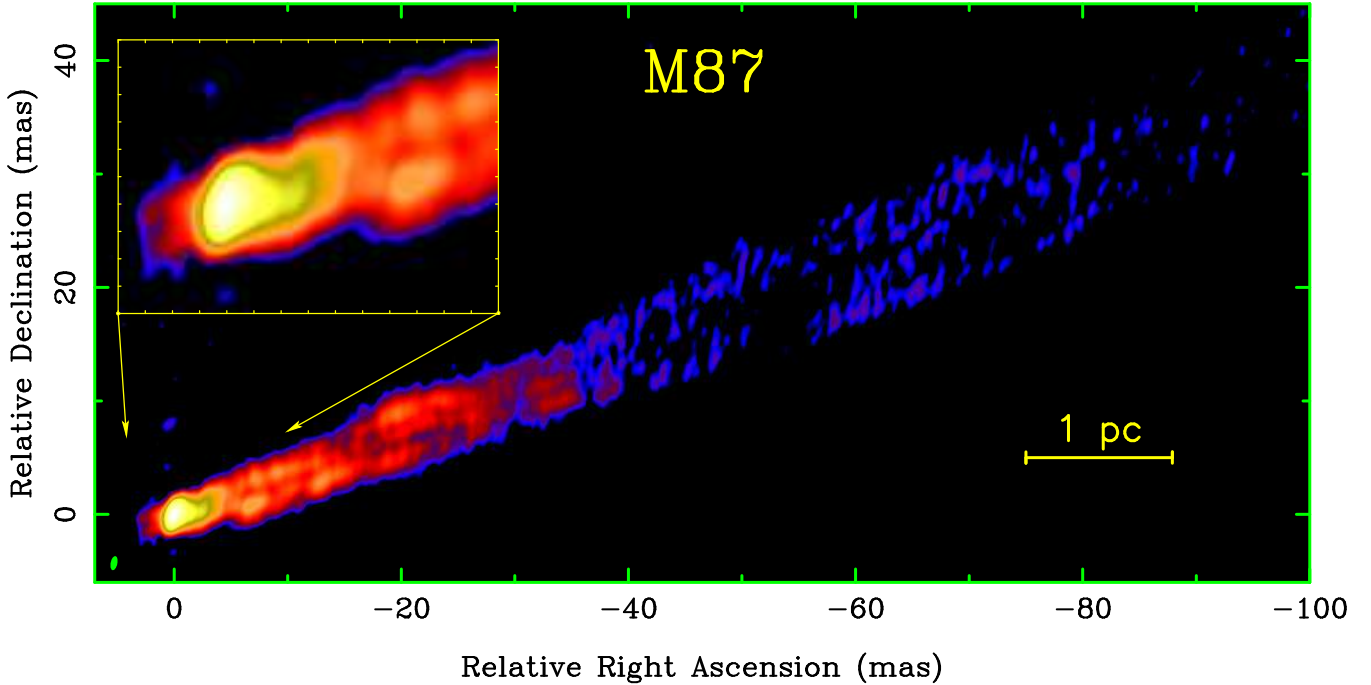}
}
\end{center}
\caption{\label{f:M87_VLBA}
VLBA 2~cm logarithmic false-color image of the M87 jet. The beam, FWHM
of 0.6 by 1.3~mas at P.A.~$=-11^\circ$ is shown in the lower left hand
corner. The peak intensity is 1.00~Jy\,beam$^{-1}$ and the off-source
rms noise 64~$\mu$Jy\,beam$^{-1}$. The corresponding dynamic range is
better than 15,000 to 1. This image was obtained from a full track in
hour angle using the VLBA along with one VLA antenna on 2000~May~8. The
faint counter-jet is seen extending to the southeast.
}
\end{figure*}

M87 has been regularly observed with the VLBA since 1995 as part of the
2~cm VLBA survey \citep{2cmPaperI} and the more recent \mbox{MOJAVE} program
\citep{LH05}. In these programs, at each epoch we observed each source
for a total of about one hour, with multiple observations spaced over a wide 
range of hour angle. For M87, we have obtained a total of 21 images
between 1995 and 2007. Typically the rms noise in each image is about
0.3~mJy\,beam$^{-1}$. As described in more detail by
\cite{2cmPaperI,2cmPaperIII} and \cite{LH05}, for each epoch, the data
were calibrated using AIPS and imaged with Difmap. The location of each
bright feature was fitted in the visibility plane.

We have supplemented these multiple epoch images using VLBA 2~cm archive
data from observations made at three epochs in 2000 (22~January, 8~May,
30~December). These later observations included one VLA antenna and were
made with full tracks in hour angle each lasting about 10 hours using
2-bit recording at  a 256~Mbps data rate, allowing us to reach the
thermal noise level  of less than 70~$\mu$Jy\,beam$^{-1}$. 

As a check on the robustness of the weak features seen in these high
dynamic range images, different authors independently made images using
Difmap and AIPS respectively and obtained very similar results. In
addition, to verify the reality of faint ``counter-feature'' reported in
\S\ref{s:structure}, one of us  generated three artificial datasets of
M87. These model datasets were based on the structure of M87 which was
observed on 2000~May~8. They used the identical ($u$,$v$)-coverage of
that epoch and contained a  comparable level of random noise.  One of
the models contained an extended counter-feature totaling nearly 40~mJy,
and two did not. All three model datasets were independently
self-calibrated and imaged by another one of us without any prior
communication about which models  contained the counter-feature.  The
image maker was able to correctly identify the data with the
counter-feature and correctly determined its flux density and extended
structure. No counter-feature was found in the derived image down to the
noise level from the other two datasets without the counter-feature. 
Given the results of these trials, and the presence of the
counter-feature in all three observed full-track VLBA images as well as
in the 20 lower sensetivity 2~cm Survey / \mbox{MOJAVE} images , we are
confident that the observed counter-feature is real and that there are
no significant artifacts in our image created by the imaging process.

\section{JET STRUCTURE }
\label{s:structure}

In Figure~\ref{f:M87_VLBA}, we show the 2~cm image constructed from
observations made with the VLBA plus 1 VLA antenna. A tapered image 
made with twice the beam size from the same data shows structure out to
nearly 0.2~arcsec.  As seen in Figure~\ref{f:M87_VLBA}, the jet appears
bifurcated, starting at about 5 mas (0.4~pc) from the core,
characteristic of a single limb brightened cylindrical or conical jet. 
The M87 jet appears to be highly collimated, with re-collimation observed
between 2~pc where the opening angle is about $16^\circ$, and 12~pc where
the opening angle is only $6^\circ$ to $7^\circ$. As discussed in
\S\ref{s:obs}, Figure~\ref{f:M87_VLBA} also shows the existence of weak
structure  extending away from the bright core toward the southeast. 
Indications of this counter feature were first suggested by 
\cite{Ly_etal04} on the basis of their 7\,mm VLBA observations. This
counter-feature also appears clearly bifurcated but is weaker than the
main jet by a factor of 10 to 15 close to the core and is only traced
out to 3.1~mas from the core. Between 3.1 and 6~mas away, this feature
is weaker than the  jet by at least a factor of 200.  

For several reasons we believe that the eastern extension may be the
counter-jet.  Based on their higher resolution 7~mm image, \cite{LWJ07}
have also detected the counter feature and have argued that the true
base of the jet cannot be offset by more than 2~mas from the bright
core; whereas we have detected the counter-feature to be at least
3.1~mas long.  Also, we note strong circular polarization at a
fractional level of $-0.5\%\pm0.1\%$ was detected at 2~cm by \cite{HL06}
coincident with the flux density peak, suggesting that this region of
the jet has an optical depth near unity \citep[e.g.,][]{J88} which is
characteristic of jet cores. Keeping in mind that there is possible ambiguity
in the VLBI image alignment, \cite{ZT03} have measured the spectral
index between 8~GHz and 12~GHz for the region with the peak intensity to
be $\alpha=0.24$ ($S\sim\nu^\alpha$) consistent with an optically thick
nucleus. Since the surface brightness of the eastern feature is more
than 200 times fainter than the one of the brightest feature which we
have identified with the core, it seems unlikely that it could be the
actual core. However, considering that no other radio jet has been
observed with comparable linear resolution, sensitivity, and dynamic
range, we cannot exclude the possibility that we are seeing the detailed
structure of the  optically thick core of the jet, and that the
counter-jet itself is not detected.

\section{JET KINEMATICS}
\label{s:kinematics}

\begin{figure}[b]
\begin{center}
\resizebox{\hsize}{!}{
   \includegraphics[trim = 0cm 0cm 0cm 0cm,angle=-90]{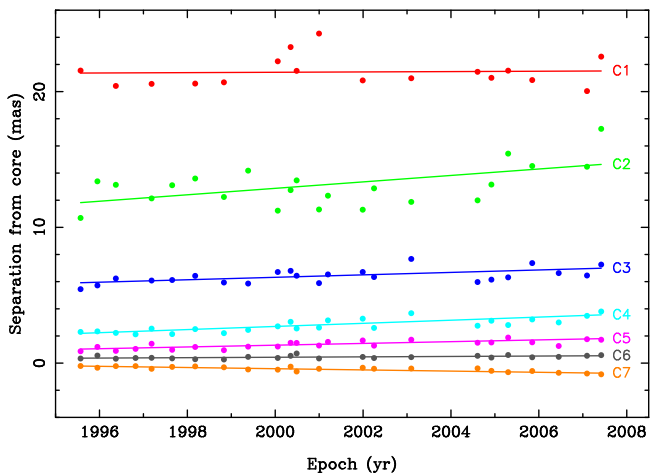}
}
\end{center}
\caption{\label{f:speed}
The location of seven separate features which were seen at multiple epochs
are plotted as a function of time.
Component C7 refers to the counter-jet.
Typical errors of the components' separation are presented
in Table~\ref{t:kinematics}.
The large scatter in the positions of the outer two components, C1
and C2, is due primarily to their large angular extent
which makes location of their centroids uncertain
in our snap-shot images of the limb-brightened jet.
}
\end{figure}

\begin{deluxetable}{lccrrr}
\tablecolumns{6}
\tabletypesize{\scriptsize}
\tablewidth{0pt}
\tablecaption{\label{t:kinematics} M87 jet components speed}
\tablehead{
\colhead{Component} & \colhead{$<R>$} &  \colhead{$\sigma_R$} & \colhead{$<\vartheta>$} & \colhead {$\mu_\mathrm{r}$}    & \colhead{$\beta_\mathrm{app}$} \\
\colhead{}          & \colhead{(mas)} &  \colhead{(mas)}      & \colhead{(deg)}         & \colhead{($\mu$as\,yr$^{-1}$)} & \colhead{}                     \\
\colhead{(1)}       & \colhead{(2)}   &  \colhead{(3)}        & \colhead{(4)}           & \colhead{(5)}                  & \colhead{(6)}
}
\startdata
C7 & 0.43 & 0.1 & 111.2 & $40\pm6$~\,& $0.010\pm0.001$ \\
C6 & 0.43 & 0.1 & 279.2 & $13\pm6$~\,& $0.003\pm0.001$ \\
C5 & 1.4  & 0.2 & 283.8 & $59\pm11$  & $0.015\pm0.003$ \\
C4 & 2.8  & 0.3 & 283.5 & $104\pm17$ & $0.026\pm0.004$ \\
C3 & 6.4  & 0.5 & 282.8 & $82\pm26$  & $0.021\pm0.007$ \\
C2 & 13   & 1.3 & 286.6 & $216\pm81$ & $0.05\pm0.02$~\,\\
C1 & 21   & 1.1 & 290.0 & $12\pm76$  & $0.00\pm0.02$~\,\\
\enddata
\tablecomments{
Col.~(1): Component identifier,
Col.~(2): Mean radial position relative to the core,
Col.~(3): Typical radial position error calculated from the scatter of
          points from the linear fit (Figure~\ref{f:speed}),
Col.~(4): Mean position angle relative to the core,
Col.~(5): Angular radial speed and $1\sigma$ uncertainty.
Positive speed means that a component is receding from the core,
negative---approaching the core,
Col.~(6): Radial speed in units of the speed of light and $1\sigma$ uncertainty.
}
\end{deluxetable}

HST observations of the M87 jet made between 1994 and 1998 by
\cite{BSM99} have suggested superluminal speeds up to about $6c$ in a
region approximately 6~arcsec (0.5~kpc) downstream of the central AGN,
consistent with what might be expected from  Doppler boosting of this
highly asymmetric jet. The feature, known as HST-1, is located 0.86
arcsec from  the base of the jet, and was more recently observed by
Chandra and HST to vary dramatically at both X-ray and optical
wavelengths \citep{H03,HCB06,CHS07}. It was found to have apparent
superluminal speeds up to $4.3\pm0.7c$ in 20~cm VLBA experiment by
\cite{CHS07}. However, within 1~arcsec of the nucleus (80~pc), using
HST, \cite{BSM99} measured much slower speeds between $0.6c$ and $0.8c$.

At radio wavelengths, \cite{BZO95} reported VLA observations which
showed a typical speed of about $0.5c$ but ranging up to $2.5c$ at
distances up to about 20~arcsec (1.6~kpc) from the nucleus.  However,
previous VLBA observations \citep{BJ95,JB95} and VSOP (VLBA to HALCA)
observations \citep{DEH06} found no evidence of motions within 5~pc of
the core, although \cite{RBJM89} reported an observed velocity of
$0.28c\pm0.08c$ for a feature located about 20~mas (1.6~pc) from the
nucleus. More recently, based on just two of five total epochs,
\cite{LWJ07} reported possible speeds between $0.25c$
and $0.4c$ for three transient features located about 3~mas (0.24~pc)
from the nucleus.

We find no evidence for motions faster than $0.07c$ within the inner 20
mas (1.6~pc). In Figure~\ref{f:speed} we plot the location of seven
separate jet features as a function of time, and we summarize their
speeds in Table~\ref{t:kinematics}. All the jet components were fit with
circular Gaussians. The fastest jet speed observed is only
$0.026\pm0.004c$, while the counter-jet feature apparently moves outward at
$0.010\pm0.001c$. The other five features appear essentially stationary
over the twelve years that they have been observed, with nominal upper
limits to their speed of about $0.07c$.
Each feature was detected at 17 to 24 separate epochs.

Given that much faster speeds, up to $6c$, have been observed in the
arcsecond scale jet, it is important to consider the possibility of
temporal under-sampling in our kinematic analysis of the milli-arcsecond
scale jet, where $6c$ would correspond to $\simeq 24$~mas\,yr$^{-1}$.
Our VLBA sampling interval ranged from 2 to 8~months, with a median of
5~months. A newly ejected, fast component could therefore conceivably
travel anywhere between 2 to 16~mas downstream in the time between
successive epochs, possibly leading to an incorrect component
cross-identification in Figure~\ref{f:speed}. However, the large number
of jet components present at each epoch implies that if they have very
fast speeds, new components would have to be emitted very frequently
from the core region, at a rate of $\sim 2$ per year. But, the flux
density and brightness temperature of the core show no indication of
such activity, but instead vary relatively smoothly over time. Second,
there are clear gaps in jet brightness between 3 and 5~mas and between 6
and 10~mas (see Figure~\ref{f:speed} and the movie in the electronic
edition of the Journal) in which jet components are never present at any
epoch~--- a highly unlikely situation given a high component ejection
rate. Third, the brightness temperatures of the individual components C1
to C7 are consistent across the epochs, which would not be expected if
our cross-identifications were incorrect. Finally, our 2~cm VLBA movie
of the jet\footnote{A movie of the jet covering the period 1999 to 2007
can be seen in electronic edition of the Astrophysical Journal. The maps
prior to 1999 have a higher noise, they were not included in the
movie. An up to date version of the movie, as well as a journal of the
observations, image and visibility FITS files can be found in our web
database
http://www.physics.purdue.edu/MOJAVE/sourcepages/1228+126.shtml}, which
relies on simple linear interpolation across the epochs, shows no sudden
jumps that would be indicative of temporal under-sampling.

We believe that this interpretation is more robust than the results of other
radio measurements which typically used only two or three epochs. We
cannot rule out the possibility of a smooth relativistic flow with no
propagating shocks or other patterns.

\section{DISCUSSION}
\label{s:discussion}

Considering that the Doppler boosted relativistic jet of M87 appears
one-sided out to kiloparsec scales, the absence of any clear motions in
the inner jet is somewhat surprising. We measure a jet to counter-jet
flux density ratio of between 10 and 15 (in the year 2000) at a position
between 0.5 and 3.1 mas from the core, and a lower limit of 200 between
3.1 to 6.0~mas away. Assuming that the jet is intrinsically
bidirectional and symmetric, and that the bulk velocity flow is the same
as the pattern motion, our maximum reliable observed speed of
$\beta_\mathrm{app}=0.026\pm0.004$ would imply a jet to counter-jet
ratio near unity for the commonly accepted viewing angle of $\sim
40^\circ$ \citep[e.g.][]{OHC89, RBJM89}. If the jet and counter-jet are
intrinsically symmetric and oriented at an angle of $30^\circ$ to
$40^\circ$ to the line of sight, the observed jet to counter-jet flux
density ratios imply that the intrinsic flow speed, $\beta$, is 0.5 to
0.6 between 0 and 3.1~mas from the core and increases to $>$0.9 beyond
3.1~mas. Also, we note that the rather small ratio of observed jet to
counter-jet speeds (Table~\ref{t:kinematics}) of less than 2.5 is not
consistent with simple relativistic beaming models. We conclude that
either the inner jet is intrinsically asymmetric, or there are no detectable
moving features within a rapidly flowing plasma.

Apparent limb brightening is predicted from analytic and numerical
modeling of relativistic jets \citep{Aloy_etal2000,Perucho_etal07} and
can  be reproduced via Kelvin-Helmholtz instability, which is
seen in extragalactic jets both at kiloparsec \citep[M87,][]{LHE03} and
parsec scales \citep[3C273,][]{LZ01}. Limb brightening can be
particularly pronounced when the jet opening angle is greater than the
beaming angle \citep{GWD06} and especially if there is a velocity
gradient across the jet \citep{GDSW07}. A two layer ``spine-sheath''
model has been suggested to explain the existence of observed strong TeV
emission from BL~Lacs with apparently slow moving radio jets
\citep{CCCG00,PE04,G04}. \cite{SO02} and \cite{GTC05} have considered a
two component jet having a fast spine which produces the gamma ray
emission by inverse Compton scattering of the radio photons from a
surrounding slow layer or sheath. Spine-sheath models have also been
discussed by \cite{B96}, \cite{SBB98}, \cite{ARW99}, \cite{LB02}, and
\cite{Cohen_etal07}. The central gap seen in our VLBA images of M87, as
well as observed TeV emission \citep[][and as discussed above]{A06}
combined with the lack of measurable motion within the inner 1.6~pc
(\S\ref{s:kinematics}) appear to support this two component model. 

The recently reported detection of strong variable TeV emission from M87
\citep{A06} also presents a problem, since unless the radiating
plasma has a large Doppler factor, energy losses due to $\gamma -
\gamma$ pair production will extinguish the gamma ray emission
\citep[e.g.,][]{DG95}.
Although we do not find any evidence for a fast
moving jet in M87 close to the central engine, the observed TeV
emission can be explained in terms of a dual layer model with a fast
inner jet and a slower moving outer layer \citep{GTC05}. In this
picture, the inner jet is beamed away from us and is thus not seen in
our VLBA images, and we only observe the slower outer layer.  However,
even the slow outer layer must move at at~least $\beta>0.5$--0.8 to
be consistent with the jet to counter-jet flux density ratio as
discussed earlier in this section.

\cite{CHS07} argue that the superluminal feature HST-1, located at a
de-projected distance of more than 100~pc down the jet, may be the
source of TeV emission in M87, rather than the base of the jet. By
analogy, they suggest that blazar activity more broadly may not, as is
widely assumed, be located close to the central engine. If this is the
case, we might expect to observe small-scale structure in HST-1 at 2~cm,
but we do not find any milliarcsecond structure stronger than 0.75~mJy
in that region in the 2000 full-track VLBA data; however our data were
taken five years before the HST-1 flaring  event in 2005.  \cite{A06}
commented that an origin of the TeV emission in HST-1 would imply an
unrealistically small opening angle for the energy source, assuming the
source was located at the base of the jet.

\acknowledgments
The VLBA is a facility of the National Science Foundation operated by
the National Radio Astronomy Observatory under cooperative agreement
with Associated Universities, Inc.  Part of this work made use of
archived VLBA and VLA data obtained by J.~Biretta, F.~Owen, W.~Junor,
and one of us (KIK). YYK is a Research Fellow of the Alexander von
Humboldt Foundation. YYK was partly supported by the Russian Foundation
for Basic Research (grant 05-02-17377). DCH was partially supported by
an award from the Research Corporation. Part of this work was done by
YYK, DCH, and MLL during their Jansky fellowship at the
NRAO. The \mbox{MOJAVE} project is supported under NSF grant AST-0406923 and a
grant from the Purdue Research Foundation. We thank T.~Cheung,
D.~Harris, A.~Lobanov, E. Ros, C.~Walker, and the \mbox{MOJAVE} team for
helpful discussions and contributions to this paper. We also appreciate
helpful comments of the anonymous referee.


{\it Facilities:} \facility{VLBA, VLA}.

\bibliographystyle{apj}

\end{document}